# Nanometric square skyrmion lattice in a centrosymmetric tetragonal magnet

Nguyen Duy Khanh[1*†], Taro Nakajima[1,2], Xiuzhen Yu[1], Shang Gao[1‡], Kiyou Shibata[1§],
Max Hirschberger[1,3], Yuichi Yamasaki[1,4,5], Hajime Sagayama[6], Hironori Nakao[6],
Licong Peng[1], Kiyomi Nakajima[1], Rina Takagi[1,3,7], Taka-hisa Arima[1,8],
Yoshinori Tokura[1,3,9] and Shinichiro Seki[1,3,5,7]*

[1] RIKEN Center for Emergent Matter Science (CEMS), Wako, Japan

[2] Institute for Solid State Physics (ISSP), The University of Tokyo, Chiba, Japan

[3] Department of Applied Physics, The University of Tokyo, Tokyo, Japan

[4] Research and Services Division of Materials Data and Integrated System (MaDIS),
National Institute for Materials Science (NIMS), Tsukuba, Japan

[5] PRESTO, Japan Science and Technology Agency (JST), Kawaguchi, Japan

[6] Institute of Materials Structure Science, High Energy Accelerator Research Organization,
Tsukuba, Ibaraki, Japan

[7] Institute of Engineering Innovation, The University of Tokyo, Tokyo, Japan

[8] Department of Advanced Materials Science, The University of Tokyo, Kashiwa, Japan

[9] Tokyo College, The University of Tokyo, Tokyo, Japan

† Current address: Institute for Solid State Physics (ISSP), The University of Tokyo, Chiba, Japan

‡ Current address: Materials Science & Technology Division and Neutron Science Division, Oak Ridge National Laboratory, Oak Ridge, TN, USA

§ Current address: Institute of Industrial Science, The University of Tokyo, Meguro, Tokyo, 153-8505, Japan

*khanh.nguyen@riken.jp; seki@ap.t.u-tokyo.ac.jp

**Magnetic skyrmions are topologically stable spin swirls with particle-like character and potentially suitable for the design of high-density information bits. While most known skyrmion systems arise in noncentrosymmetric systems with Dzyaloshinskii-Moriya interaction, also centrosymmetric magnets with a triangular lattice can give rise to skyrmion formation, with geometrically-frustrated lattice being considered essential in this case. Until today, it remains an open question if skyrmions can also exist in the absence of both geometrically-frustrated lattice and inversion symmetry breaking. Here, we discover a square skyrmion lattice state with 1.9 nm diameter skyrmions in the centrosymmetric tetragonal magnet $GdRu_2Si_2$ without geometrically-frustrated lattice by means of resonant X-ray scattering and Lorentz transmission electron microscopy experiments. A plausible origin of the observed skyrmion formation is four-spin interactions mediated by itinerant electrons in the presence of easy-axis anisotropy. Our results suggest that rare-earth intermetallics with highly-symmetric crystal lattices may ubiquitously host nanometric skyrmions of exotic origins.**

Ordering of magnetic moments in non-collinear or non-coplanar textures has been known as a source of plentiful and intriguing phenomena[1]. Among them, the magnetic skyrmion generally appears as a vortex-like swirling spin texture with non-zero integer skyrmion number $n_{sk}$ defined by

$$n_{sk} = \frac{1}{4\pi} \int \boldsymbol{n} \cdot \left( \frac{\partial \boldsymbol{n}}{\partial x} \times \frac{\partial \boldsymbol{n}}{\partial y} \right) dxdy, \qquad (1)$$

which represents how many times the spin directions wrap a unit sphere[2-4]. Here, the integral is taken over the two-dimensional magnetic unit cell, and $\boldsymbol{n}(\boldsymbol{r}) = \boldsymbol{m}(\boldsymbol{r})/|\boldsymbol{m}(\boldsymbol{r})|$ is the unit vector pointing along the local magnetic moment $\boldsymbol{m}(\boldsymbol{r})$. This indicates that a magnetic skyrmion has the character of a countable particle. Its size typically ranges from a few to hundreds of nanometers. In metallic materials, the skyrmion motion can be efficiently driven by the electric current through spin transfer torque, and the interplay between conduction electron and the skyrmion spin texture via quantum Berry phase also leads to unique transport phenomena such as the topological Hall effect[5,6]. Such a stable particle nature, small size, and electric controllability highlight the magnetic skyrmion as a potential new information carrier for high-density magnetic storage devices[7].

Previously, the emergence of magnetic skyrmions has mostly been reported for a series of materials with noncentrosymmetric structures, where the inherent spin-twisting interaction termed Dzyaloshinskii-Moriya (DM) interaction plays a crucial role in the skyrmion formation[8-10]. In such systems, helical spin texture, $\boldsymbol{m}(\mathbf{r}) = \boldsymbol{m_Q} \exp[i\boldsymbol{Q} \cdot \boldsymbol{r}] + c.c.$, (with $\boldsymbol{m_Q}$ being a complex number) characterized by single magnetic modulation vector $\boldsymbol{Q}$ (i.e. single-$\boldsymbol{Q}$ state) is realized for magnetic field $B = 0$. The application of external magnetic field stabilizes the triangular lattice of skyrmions, which can be approximately described by $\boldsymbol{m}(\mathbf{r}) = (0,0, m_0^z) + \sum_{v=1,2,3}(\boldsymbol{m_{Q_v}} \exp[i\boldsymbol{Q_v} \cdot \boldsymbol{r}] + c.c)$ + higher harmonics with three distinctive fundamental modulation vectors $\boldsymbol{Q_v}$ (triple-$\boldsymbol{Q}$ state). Therein, the internal spin texture of the skyrmion is determined by the crystallographic symmetry of the target system, and the observation of Bloch type[11-15], Neel type[16,17], and anti-vortex type skyrmion (or antiskyrmion)[18] has been reported for chiral, polar, and $D_{2d}$ symmetry of bulk magnets, respectively. On the other hand, recent theoretical studies suggest that skyrmions can be stabilized even in centrosymmetric systems by considering different microscopic mechanisms[19]. For example, geometrical frustration of short-range exchange interactions on triangular lattice is predicted to stabilize a hexagonal lattice of skyrmions[20-22]. Another potential mechanism is the interplay of Ruderman-Kittel-Kasuya-Yosida (RKKY) and four-spin interactions associated with s-d or s-f coupling mediated by itinerant electrons, which is expected to favor a multiple-$\boldsymbol{Q}$ skyrmion lattice state for highly-symmetric (such as hexagonal or tetragonal) crystal lattice system[23-28]. In the latter scenario, the wave vector of magnetic modulation is governed by the long-range RKKY interaction, which can cause a kind of magnetic frustration.

Here, an important challenge is the search for an appropriate material system to experimentally realize these situations. Very recently, it has been discovered that the centrosymmetric triangular-lattice magnet $Gd_2PdSi_3$ hosts a hexagonal lattice of skyrmions, which is characterized by the small skyrmion diameter $\lambda_{sk} \sim 2.4$ nm with an associated giant topological Hall effect[29]. Similar skyrmion formation has also been reported for centrosymmetric $Gd_3Ru_4Al_{12}$ with breathing Kagomé network[30]. For these compounds, the existence of geometrically-frustrated lattice is considered to be the key for the skyrmion formation. Nevertheless, the competition of long-range magnetic interactions mediated by itinerant electrons can be allowed for any types of crystal lattice, in principle. Therefore, it remains an important question whether similar skyrmion formation in centrosymmetric magnets without geometrically-frustrated lattice is possible or not. In this study, we focus on the centrosymmetric tetragonal magnet GdRu$_2$Si$_2$ without geometrically-frustrated lattice, and investigated its magnetic structure in detail. We found the emergence of the double-$Q$ square skyrmion lattice state in an out-of-plane magnetic field. The observed skyrmion diameter is as small as 1.9 nm, which is the smallest among ever reported for single-component bulk materials. Our present results demonstrate that skyrmion lattice can be formed even without geometrically-frustrated lattice nor inversion symmetry breaking, and suggest that rare-earth intermetallics with highly symmetric crystal lattice in general can be a promising material platform to realize nanometric skyrmions of exotic origins.

**Magnetic phase diagram for *B* // [001]**

GdRu$_2$Si$_2$ belongs to the family of *RM$_2$X$_2$* compounds crystallized in ThCr$_2$Si$_2$-type structure with centrosymmetric tetragonal space group *I4/mmm* (*R*: rare-earth element, *M*: 3d, 4d or 5d element and *X*: Si or Ge)[31]. The crystal structure consists of alternate stacking of square lattice Gd layers and Ru$_2$Si$_2$ layers as presented in Fig. 1a, where the magnetism is governed by Gd$^{3+}$ (*S* = 7/2, *L* = 0) ions with Heisenberg magnetic moment. According to previous reports, this compound hosts incommensurate magnetic order below $T_N \approx$ 46 K[31,32], with magnetic modulation vector $Q$ = (0.22, 0, 0) confined within the tetragonal basal plane[33]. Application of a magnetic field *B* along the [001] axis induces several magnetic phase transitions as summarized in Fig. 1b, while the detailed magnetic structure in each phase has not been identified[31-34].

Figures 1f-h and j-l indicate magnetic field dependence of magnetization *M*, longitudinal resistivity $\rho_{xx}$, and Hall resistivity $\rho_{yx}$ measured at 5 K for *B* // [001]. The magnetization profile shows clear step-like anomalies at 2.0 T and 2.4 T, before reaching the saturated ferromagnetic state with *M* ~ 7$\mu_B$/Gd$^{3+}$ at 10 T. These magnetic transitions are accompanied by large changes in $\rho_{xx}$ and $\rho_{yx}$, suggesting close correlation between magnetic orders and transport properties[31]. Figures 1b and i summarize the *B-T* magnetic phase diagram of GdRu$_2$Si$_2$ for *B* // [001], where three distinctive magnetic phases (Phases I, II and III) can be identified below 20 K.

**Magnetic structure analysis**

To resolve the detailed magnetic structure in each phase, we performed magnetic X-ray scattering experiments in resonance (RXS) with the Gd *L$_2$* edge at 5 K. By exploring the magnetic satellite peaks around the fundamental Bragg reflection indexed as (4, 0, 0) $\pm$ $Q$, the magnetic modulation vector $Q$ in each phase has been investigated directly. In Figs. 2a,d,g, the scattering profiles along (4-$\delta$, 0, 0) measured at 0 T, 2.1 T, and 3 T (corresponding to Phases I, II, and III, respectively) are plotted, where the existence of the magnetic modulation vector $Q_1$ = ($q$, 0, 0) with $q \sim 0.22$ is clearly identified in all three phases. Because of the tetragonal symmetry of the crystal structure, the appearance of an equivalent magnetic modulation vector $Q_2$ = (0, $q$, 0) is expected, which has also been identified in the scattering profiles along (4, -$\delta$, 0) shown in Figs. 2b,e,h. In Figs. 2j-l, magnetic field dependence of magnetization, wave number $q$ of magnetic modulation, and integrated intensity of the magnetic peak are plotted. Upon the magnetic phase transitions at 2.0 T and 2.4 T, both $q$-value and peak intensity show clear step-like anomalies.

Here, the simultaneous appearance of $Q_1$ = ($q$, 0, 0) and $Q_2$ = (0, $q$, 0) can be interpreted as either a double-$Q$ magnetic state or multiple-domains of a single-$Q$ magnetic state. One of the most effective ways to distinguish these two possibilities is the identification of additional $Q_1$ + $Q_2$ modulations[35], which are allowed to appear only for the former double-$Q$ magnetic state. In Figs. 2c,f,i, the corresponding scattering profiles along (4-$\delta$, -$\delta$, 0) are plotted, where the $Q_1$ + $Q_2$ modulation is identified only in the intermediate Phase II, but not in Phase I or III. Such a feature can be more directly confirmed in the magnetic field dependence of the integrated intensity for the $Q_1$ + $Q_2$ magnetic peak (Fig. 2m), in which the appearance of the $Q_1$ + $Q_2$ modulation is clearly correlated with the transition into Phase II. On the basis of these results, we conclude that Phase II is a double-$Q$ magnetic state.

**Spin texture of magnetic phases**

Next, to investigate the spin orientation in each magnetic phase, we analyze the polarization of the scattered X-ray. The experimental geometry is illustrated in Figs. 3a and b, in which the magnetic satellite peaks at the (4, −2, 0) $\pm$ $Q$ position are investigated. Here, the incident beam is always polarized parallel to the scattering plane ($\pi$-polarized). The scattered beam may include two polarization components: parallel ($\pi'$) and perpendicular ($\sigma'$) polarizations. The intensities of the two components of the scattered beam ($I_{\pi\text{-}\pi'}$ and $I_{\pi\text{-}\sigma'}$, respectively) are measured separately. When the magnetic structure contains the modulated spin component ($\boldsymbol{m_Q} \exp[i\boldsymbol{Q} \cdot \boldsymbol{r}] + c.c.$), the corresponding magnetic scattering intensity is given by

$$I \propto |(\boldsymbol{e}_i \times \boldsymbol{e}_f) \cdot \boldsymbol{m}_Q|^2 \quad (2)$$

with $\boldsymbol{e}_i$ and $\boldsymbol{e}_f$ representing the polarization vectors of incident and scattered beams, respectively[36]. Since $\boldsymbol{e}_i$ is almost parallel to the [010] axis in the present experimental geometry, $I_{\pi\text{-}\pi'}$ and $I_{\pi\text{-}\sigma'}$ should mainly reflect the [001] and [100] components of $\boldsymbol{m}_Q$. On the basis of the above discussion, we first investigate the magnetic structure at 0 T, corresponding to Phase I. In Fig. 3c, the existence of $I_{\pi\text{-}\pi'}$ (absence of $I_{\pi\text{-}\sigma'}$) intensity indicates the existence of [001] (absence of [100]) modulated spin component for $\boldsymbol{Q}_1 = (q, 0, 0)$. Likewise, Fig. 3e demonstrates the coexistence of both [001] and [100] modulating spin components for $\boldsymbol{Q}_2 = (0, q, 0)$. Note that there appears an addition peak at $\delta \sim 0.226$ in (4, -2+$\delta$, 0) scan having the same selection rule as Phase III, whose origin is discussed in the Supplementary Note II. These data suggest that spin texture in the Phase I can be approximately described by the screw structure, in which neighboring magnetic moments rotate within a plane normal to the magnetic modulation vector $\boldsymbol{Q}$ as shown in Figs. 3g and h. Similar measurements have also been performed for Phase II at 2.1 T (Figs. 3d and f), where the manner of existence / absence of $I_{\pi\text{-}\pi'}$ and $I_{\pi\text{-}\sigma'}$ intensities for $\boldsymbol{Q}_1$ and $\boldsymbol{Q}_2$ turned out to be the same as Phase I (see Supplementary Note III). This means that the double-$\boldsymbol{Q}$ magnetic order in Phase II is roughly represented by a superposition of two screw spin textures with their magnetic modulation vectors orthogonal to each other. By assuming the fixed amplitude of local magnetic moment, the spin texture $\boldsymbol{m}(\boldsymbol{r})$ is approximately described by

$$\boldsymbol{m}(\boldsymbol{r}) = \boldsymbol{S}(\boldsymbol{r})/|\boldsymbol{S}(\boldsymbol{r})| \quad (3)$$

$$\boldsymbol{S}(\boldsymbol{r}) = \{(0, -i, 1) \exp[i\boldsymbol{Q}_1 \cdot \boldsymbol{r}] + (i, 0, 1) \exp[i\boldsymbol{Q}_2 \cdot \boldsymbol{r}] + (0, 0, S_z)\} + c.c. \quad (4)$$

Here, $S_z$ approximately scales with the *B*-induced out-of-plane uniform magnetization component. The normalization process leads to the emergence of higher harmonic terms in $\boldsymbol{m}(\boldsymbol{r})$ with the wave vectors such as $2\boldsymbol{Q}_1$, $2\boldsymbol{Q}_2$ and $\boldsymbol{Q}_1 + \boldsymbol{Q}_2$, whose existence has also been confirmed experimentally (See Fig. 2f, Supplementary Fig. 6, and the related discussion in Supplementary Note V).

Figure 1d describes the spin texture given by Eq. (3), which is characterized by the integer skyrmion number $n_{sk} = -1$ per magnetic unit cell according to Eq. (1) and can be considered as the square lattice of magnetic skyrmions (See Supplementary Fig. 4 and Supplementary Note IV). Similar polarization analysis has also been performed for the Phase III, whose magnetic structure is displayed in Fig. 1e. (See Supplementary Note II for the detail). In general, the SkL phase is separated by first-order phase-transition boundaries with other topologically-trivial magnetic phases[21,25,29,30], which is consistent with the observed *M-B* profile with two discontinuous magnetization steps with clear hysteresis (Fig. 2j).

Note that conduction electrons interacting with the skyrmion spin texture gain an extra Berry phase acting on the conduction electrons as an emergent magnetic field, which causes an additional contribution to the Hall resistivity proportional to the skyrmion density, i.e. topological Hall effect (THE). As shown in Figs. 1h and l, an enhancement of $|\rho_{yx}|$ is identified in Phase II (i.e. in the SkL phase), which may contain the possible contribution from the THE. Reflecting the strong correlation between magnetism and transport properties, GdRu$_2$Si$_2$ also shows non-monotonous magnetoresistance and large anomalous Hall effect (Figs. 1g and h). In such a situation, the interpretation of transport properties is not straightforward. Our tentative analysis in terms of THE is provided in Supplementary Note VI, while the quantitative evaluation remains a future challenge.

**Real space imaging of square skyrmion lattice**

Figure 4a indicates the under-focused Lorentz transmission electron microscopy (L-TEM)image obtained at 1.95 T and 8 K (i.e. in Phase II), with a Fourier transform pattern in Figs. 4b and c. Here, the peaks marked by white circles are related to the crystal structure (whose corresponding Miller indices are also indicated), and the four-fold peaks marked by yellow circles (corresponding to magnetic modulation vector $Q_1$ = (0.22, 0, 0) and $Q_2$ = (0, 0.22, 0)) are related to magnetic structure. These results are in good agreement with the RXS results. To filter undesired background noise, we pick up the circled area (i.e. reflections related to crystalline and magnetic structures) in Fig. 4b and reversed the Fourier transformation. The result is shown in Fig. 4e, which clearly visualizes the square-lattice-like magnetic super-structure with a period of 1.9 nm formed on the atomic lattice. Figures 4f and g are filtered images obtained from the under-focused and over-focused L-TEM images, respectively, based only on the magnetic reflections circled in Fig. 4c. Both under-focused and over-focused images show square-lattice-like pattern but with opposite black/white contrast. By performing the TIE (transport-of-intensity equation) analysis[15] based on these two images, the spatial distribution of in-plane local magnetization is deduced as shown in Fig. 4h. We can confirm the square lattice of vortex-like spin arrangements, which is revealed to be consistent with the spin texture in Fig. 1d representing the square skyrmion lattice described by Eq. (3). Above $T_c$, all the magnetic contrast and the corresponding magnetic reflections disappear as shown in Fig. 4d.

## Microscopic origins of square skyrmion lattice

Theoretically, several distinct mechanisms for the emergence of multiple-$Q$ spin textures have been proposed, such as (1) Dzyaloshinskii-Moriya interaction in noncentrosymmetric magnets[2,3,8], (2) competition of short-ranged exchange interactions in geometrically-frustrated magnets[20-22], and (3) interplay of RKKY and four-spin interactions related to the coupling between conduction electron and localized moment in itinerant magnets with highly-symmetric lattice[22-26]. Here, the mechanism (2) and (3) are associated with magnetic frustration in a broad sense, and the magnetic modulation is generated by the sign-alternating competing spin exchange interactions in the both cases. In the latter case for itinerant magnets, the presence or absence of magnetic frustration is not directly linked to the underlying crystal lattice geometry.

Since the crystal structure of GdRu$_2$Si$_2$ is centrosymmetric, the contribution from the DM interaction is not relevant in the present case. Instead, the family of $R$Ru$_2$Si$_2$ systems commonly shows incommensurate magnetic modulation along in-plane directions, which has been discussed in terms of $(\boldsymbol{m_Q} \cdot \boldsymbol{m_{-Q}})$-type RKKY interactions reflecting Fermi surfaces properties[31]. In this sense, the present formation of a square skyrmion lattice can potentially be ascribed to the four-spin interaction mediated by itinerant electrons[23-28]. This mechanism has originally been proposed for Fe atomic monolayer on Ir(111) surface, where the formation of a double-$Q$ square skyrmion lattice has been confirmed by spin-polarized scanning tunneling microscopy experiments[26]. A similar mechanism has also been discussed to describe the coplanar triple-$Q$ magnetic order in the hexagonal magnet Y$_3$Co$_8$Sn$_4$[27]. Here, the four-spin interaction is generally described as $\sum K_{ijkl}[(\boldsymbol{m}_i \cdot \boldsymbol{m}_j)(\boldsymbol{m}_k \cdot \boldsymbol{m}_l)]$ among four magnetic sites $i,j,k,l$ in real space or $\sum \widetilde{K}_{Q_1 Q_2 Q_3 Q_4}(\boldsymbol{m}_{Q_1} \cdot \boldsymbol{m}_{Q_2})(\boldsymbol{m}_{Q_3} \cdot \boldsymbol{m}_{Q_4}) \delta(\boldsymbol{Q}_1 + \boldsymbol{Q}_2 + \boldsymbol{Q}_3 + \boldsymbol{Q}_4)$ among the wave numbers $\boldsymbol{Q}_v$ ($v = 1,2,3,4$) giving multiple maxima in the bare susceptibility in reciprocal space, which lifts the degeneracy between multiple-$Q$ and single-$Q$ magnetic orders. The present discovery of a square skyrmion lattice in GdRu$_2$Si$_2$ suggests that a similar mechanism may also promote the SkL formation in single-component bulk materials. In this case, the tetragonal symmetry of the underlying crystal lattice allows the existence of multiple-number of equivalent magnetic modulation vectors $\boldsymbol{Q}_1$ and $\boldsymbol{Q}_2$ determined by RKKY interaction, and the additional contribution from the four-spin interaction stabilizes the double-$Q$ square skyrmion lattice orders. Recently, the tetragonal itinerant magnet CeAuSb$_2$ was reported to host a double-$Q$ collinear spin density wave state under $B$ // [001], but this material is characterized by strong Ising magnetic anisotropy and skyrmion formation is not allowed[37]. In contrast, magnetism in GdRu$_2$Si$_2$ is dominated by Gd$^{3+}$ Heisenberg spins, which should be another important factor for the present skyrmion formation. Note that GdRu$_2$Si$_2$ is characterized by moderate amplitude of easy axis anisotropy[32], which can also provide the effective four-spin interaction and stabilize the multiple-$Q$ order[19,21]. Such an interplay among the magnetic interactions mediated by itinerant electrons and the easy-axis magnetic anisotropy is perhaps responsible for the present SkL formation.

**Conclusions**

In the case of DM-induced skyrmions, a close-packed triple-$Q$ hexagonal SkL is usually favored irrespective of the underlying crystal symmetry[9-11,17,18], and the double-$Q$ square SkL is rather exceptional[13,15,38,39]. The present observation of square SkL in tetragonal GdRu$_2$Si$_2$ suggests the strong correlation between lattice symmetries of SkL and crystal structure, which can be a unique feature of centrosymmetric rare-earth intermetallics.

Interestingly, GdRu$_2$Si$_2$ has the same crystal structure as the 122-type Fe-based superconductors, and this family of materials allows wide variety of chemical parameter tuning. Because of the lack of geometrically-frustrated lattice and inversion symmetry breaking, the present $RM_2X_2$ system with its centrosymmetric tetragonal lattice may offer an ideal material platform to explore the novel skyrmion formation mechanism potentially mediated by itinerant electrons. The small skyrmion size and spin-helicity degree of freedom are another advantage of GdRu$_2$Si$_2$, and further systematic investigation of the detailed magnetic structure as well as its relationship with the electronic structure and emergent electromagnetic responses for this material family will be essential. Our results establish that skyrmions can be stabilized without the need for geometrically-frustrated lattice or inversion symmetry breaking, which suggests a new route for the design of nanometric topological spin textures in single-component systems. The further experimental identification of isolated skyrmion particle[21] in such systems would be an important challenge for the potential application.

**Acknowledgement**

We would like to thank T. Kurumaji, N. Nagaosa, R. Arita, K. Ishizaka, T. Hanaguri, Y. Motome, S. Hayami, Y. Yasui, C. J. Butler, T. Koretsune, T. Nomoto, Y. Ohigashi and A. Kikkawa for enlightening discussions and experimental helps. RXS measurements were performed under the approval of the Proposals No. 2018G570 at the Institute of Material Structure Science, High Energy Accelerator Research Organization (KEK). This work was partly supported by Grants-In-Aid for Scientific Research Scientific Research (A) (Grant No. 18H03685 (S.S) and 19H00660 (X.Z.Y)) and Grant-in-Aid for Scientific Research on Innovative Area, "Nano Spin Conversion Science" (Grant No.17H05186) from JSPS, and PRESTO (Grant No. JPMJPR18L5) from JST. M.H. was supported as a JSPS International Research Fellow (18F18804). R.T was supported by the Murata Science Foundation.

**Author contributions**

N.D.K, S.S, T.A and Y.T conceived the project. N.D.K grew single crystals and characterized magnetic and transport properties with the assistance of R.T and M.H. T.N, S.G and N.D.K carried out the RXS measurement with the assistance of K.S, Y.Y, H.S and H.N. X.Z.Y performed L-TEM observations and L.P and K.N prepared TEM samples. N.D.K and S.S wrote the manuscript. All the authors discussed the results and commented on the manuscript.

**Competing Interests**

The authors declare no competing interests.

**Methods**

**Crystal Growth**. Single crystals of GdRu$_2$Si$_2$ were grown by the optical floating zone method. Poly-crystals were prepared by the arc-melt technique from pieces of high quality elements Gd (3N), Ru (3N) and Si (5N) using a water cooled copper crucible under Ar atmosphere. The obtained crystals were characterized by powder X-ray diffraction, which confirmed the purity of the sample. Crystal orientations were determined using the back reflection X-ray Laue photography method.

**Magnetic and electrical transport property measurements**. Measurements of magnetic susceptibility (Supplementary Note I) and magnetization were performed using a superconducting quantum interference device magnetometer (MPSM 3, Quantum Design). To characterize electrical transport properties, the rectangular bar-shaped samples were used with silver paste as an electrode. Measurements of $\rho_{xx}$ and $\rho_{yx}$ were performed using the AC-transport option in a Physical Property Measurement System (PPMS). To avoid the possible discrepancy due to shape anisotropy caused by the demagnetization effect, the same crystal was used for the measurements of magnetic and electrical transport properties.

**Resonant X-ray Scattering (RXS) measurement.** RXS measurements were performed on BL-3A, Photon Factory, KEK, Japan. The photon energy of the incident X-ray was turned near the Gd $L_2$ absorption edge (~7.935 keV) with a Si (111) double-crystal monochromator. A single crystal of GdRu$_2$Si$_2$ with flat (100) plane (dimensions of 2.3×1.8×0.5 (mm)) was attached on an Al plate by varnish and loaded into a vertical-field superconducting magnet with the ($h$, $k$, 0) scattering plane (Supplementary Note VII and VIII). In this configuration, the scattering plane is perpendicular to the [001] axis. The incident X-ray was linearly polarized parallel to the scattering plane ($\pi$). The 006 reflection of a pyrolytic graphite (PG) plate was used to analyze the polarization of the scattered X-ray. The $2\theta$ angle for the analyzer at Gd $L_2$ edge was 88°. The $\sigma$' (polarized perpendicularly to the scattering plane) and $\pi$' (polarized parallel to the scattering plane) components can be selected by rotating the PG plate about the scattered beam. For the observed $Q_1 + Q_2$ reflection (in Fig. 2), the possibility of double scattering can be excluded because the present $Q_1 + Q_2$ reflection is observed only in Phase II. In addition, the penetration depth of X-ray is much smaller than that of neutron, which prevents the emergence of the double scattering process.

**Lorentz-TEM measurement.** Lorentz TEM provides information of the in-plane magnetization of the sample. The bright or dark contrast in the defocused Lorentz TEM images reflects the convergence or divergence of the deflected electron beam induced by Lorentz force, corresponding to the direction and magnitude of in-plane magnetizations. The under-focused and over-focused Lorentz TEM images should provide the reversed contrast. Accordingly, magnetic skyrmions with clockwise/counterclockwise helicity can be observed as bright/dark dots on the under-focused/over-focused image plane[40]. To examine the modulated magnetic structure in the present magnet, we thinned the bulk sample with $Ar^+$ milling after mechanical polishing and then put the thin sample on a double-tilt He cooling sample holder with a temperature-controller (Gatan ULTDT) to control the sample temperature from 6 K to 70 K, which is attached to a commercial transmission electron microscope (JEOL, JEM2800). Lorentz TEM measurements were performed under a normal magnetic field of 1.95 T. The field map for SkL was obtained by analyzing defocused Lorentz TEM images using a commercial software package Qpt based on the transport-of-intensity equation (TIE) [41].

**References for the Methods section**

**Data Availability**

The data that support the findings of this study are available from the corresponding authors upon reasonable request.

**Additional Information**

**Supplementary information** is available in the online version of the paper.

**Reprints and permission information** is available online at www.nature.com/reprints.

**Correspondence and requests for materials** should be addressed to N. D. K or S. S.

**Figures and captions.**

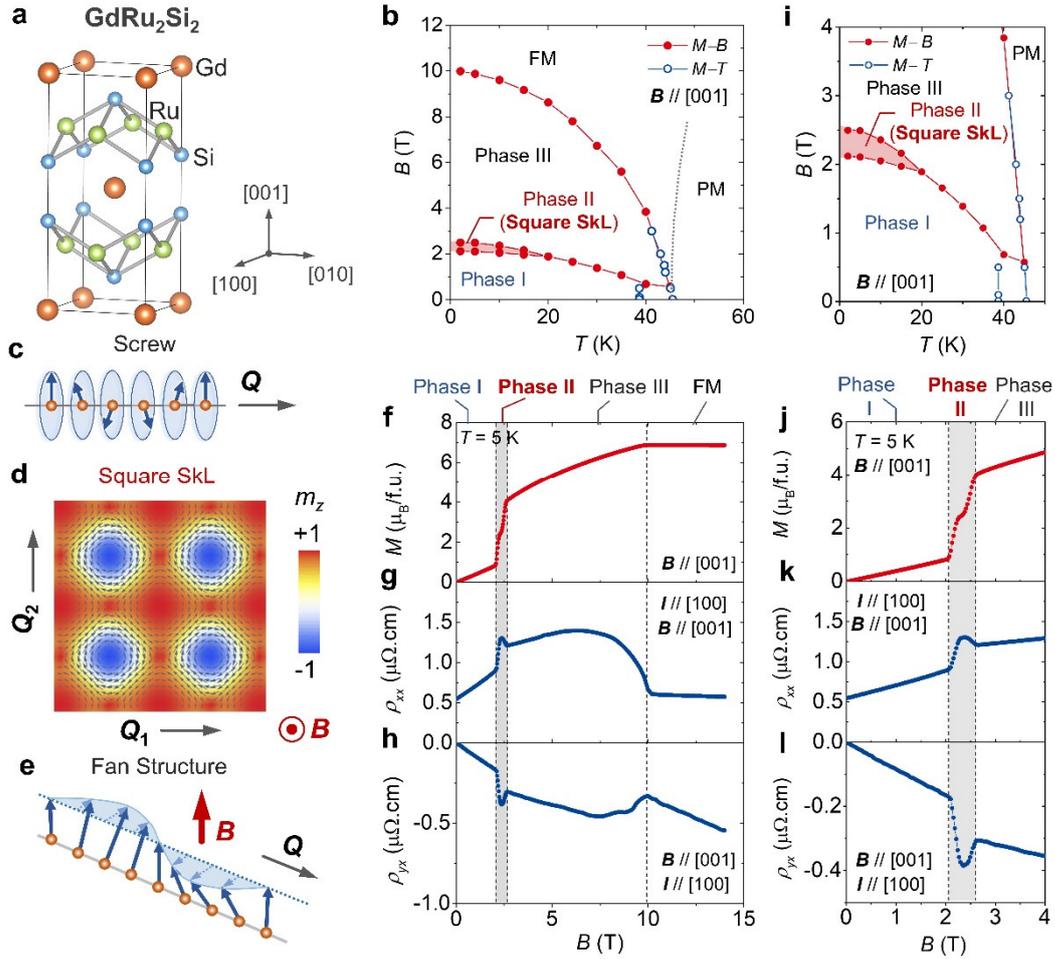

**Fig. 1 | Crystal structure, magnetic structure and magnetic phase diagram of GdRu$_2$Si$_2$. a**, Crystal structure of GdRu$_2$Si$_2$. **b**, Magnetic field ($B$) – temperature ($T$) phase diagram for $B$ // [001] based on the magnetization measurements. Multiple magnetic phases can be clearly identified, including (modulated) screw (Phase I), double-$Q$ square SkL (Phase II), and fan structure (Phase III). **c**, Schematic illustration of the screw magnetic structure in Phase I. **d**, Square SkL in Phase II described by Eq. (3), i.e. the superposition of two screw spin structures with orthogonally arranged magnetic modulation vectors $Q_1$ and $Q_2$. **e**, Fan structure in Phase III. **f-h**, Magnetic field dependence of magnetization $M$, longitudinal resistivity $\rho_{xx}$ and Hall resistivity $\rho_{yx}$ for $B$ // [001] and $I$ // [100]. (**i**) and (**j-l**) show the expanded view of data in (**b**) and (**f-h**).

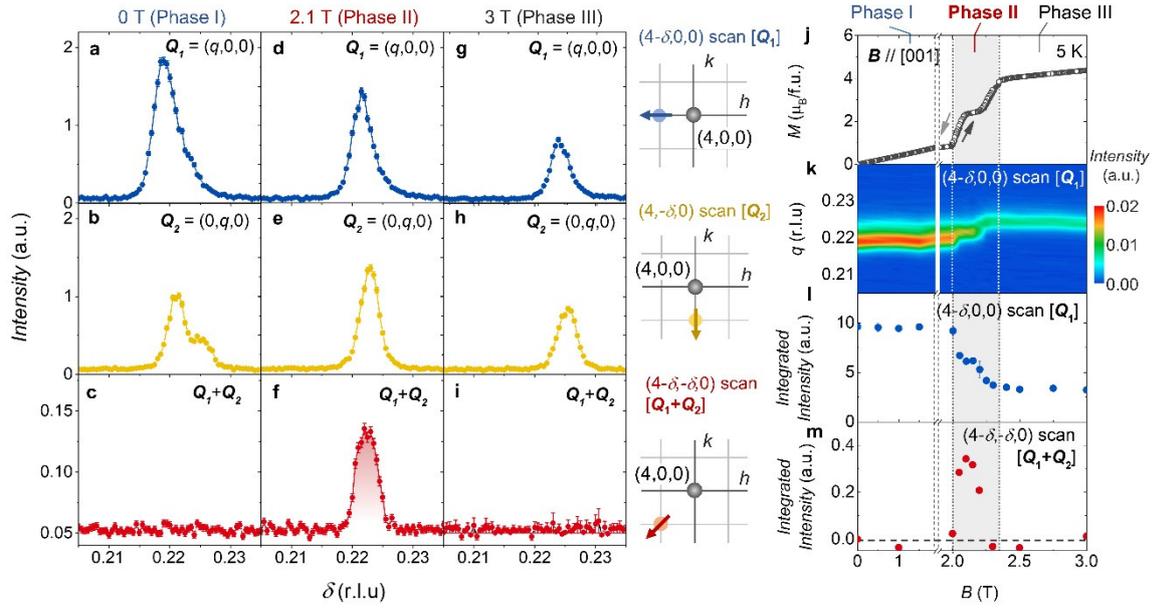

**Fig. 2 | Resonant X-ray scattering results for GdRu$_2$Si$_2$. a-c**, Line scan profiles for (4-$\delta$, 0, 0), (4, -$\delta$, 0) and (4-$\delta$, -$\delta$, 0) scans, corresponding to the **Q$_1$**, **Q$_2$**, and **Q$_1$** + **Q$_2$** magnetic satellite peaks around the fundamental Bragg spot (4, 0, 0) in Phase I ($B$ = 0 T), respectively. The corresponding data in Phase II ($B$ = 2.1 T) and Phase III ($B$ = 3 T) are plotted in (**d-f**) and (**g-i**), respectively. The magnetic field is applied along [001] axis. The line scan direction is depicted on the right panel of each figure. **j-m**, Magnetic field dependence of magnetization $M$, wave number $q$, integrated intensity of **Q$_1$** reflection and integrated intensity of **Q$_1$** + **Q$_2$** reflection. Error bars indicate one standard deviation.

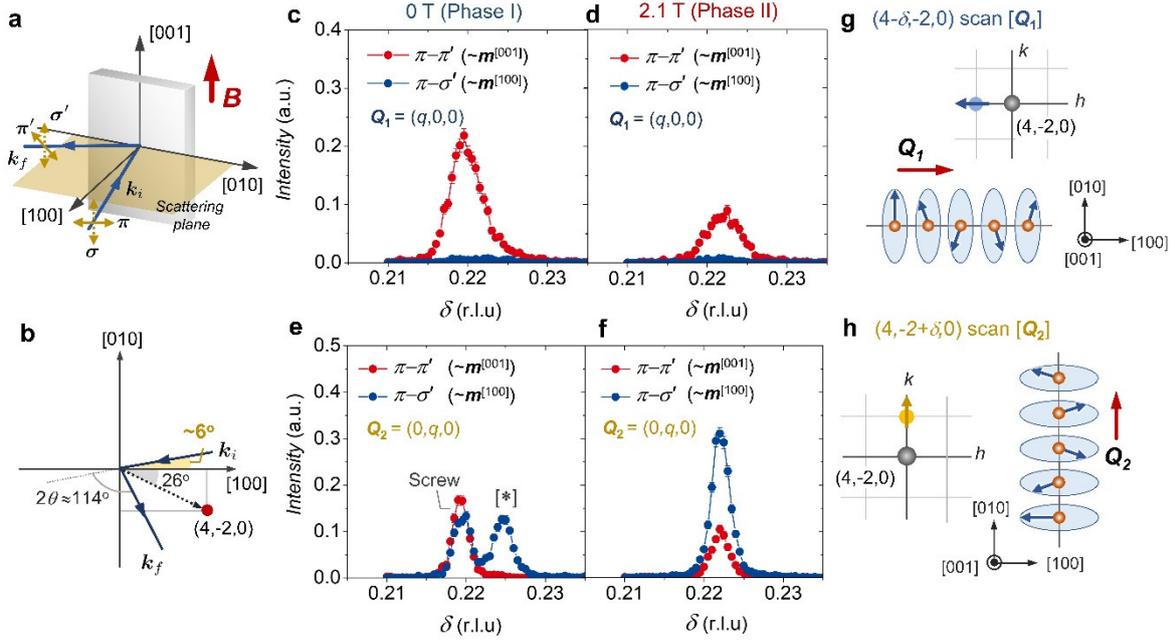

**Fig. 3 | Polarization analysis of resonant X-ray scattering (RXS) profiles in Phase I and Phase II**. **a**, Experimental setup for RXS measurement. Satellite peaks around the fundamental Bragg spot (4, -2, 0) are investigated. Scattering plane lies perpendicular to the [001] axis. $\pi$ ($\pi'$) and $\sigma$ ($\sigma'$) represent polarization direction of incident (scattered) x-ray. **b**, The measurement configuration projected along the [001] direction. In this setup, the propagation vector $k_i$ of incident x-ray is almost parallel to the [100] axis. **c,e**, Line profiles for (4-$\delta$, -2, 0) and (4, -2+$\delta$, 0) scans, corresponding to $Q_1$ and $Q_2$, measured at 0T (Phase I). **d,f**, Corresponding line profiles at 2.1 T (Phase II). As discussed in the main text, the intensities of ($\pi$-$\pi'$) and ($\pi$-$\sigma'$) channels mainly reflect the [001] and [100] components of $m_Q$, respectively. **g,h**, Line scan direction in reciprocal space, as well as a real-space illustration of modulated spin components belonging to each magnetic modulation vector. In (**e**), an additional peak (highlighted by the symbol \*) is observed at $\delta \sim 0.226$, whose origin is discussed in the Supplementary Note II. Error bars indicate one standard deviation.

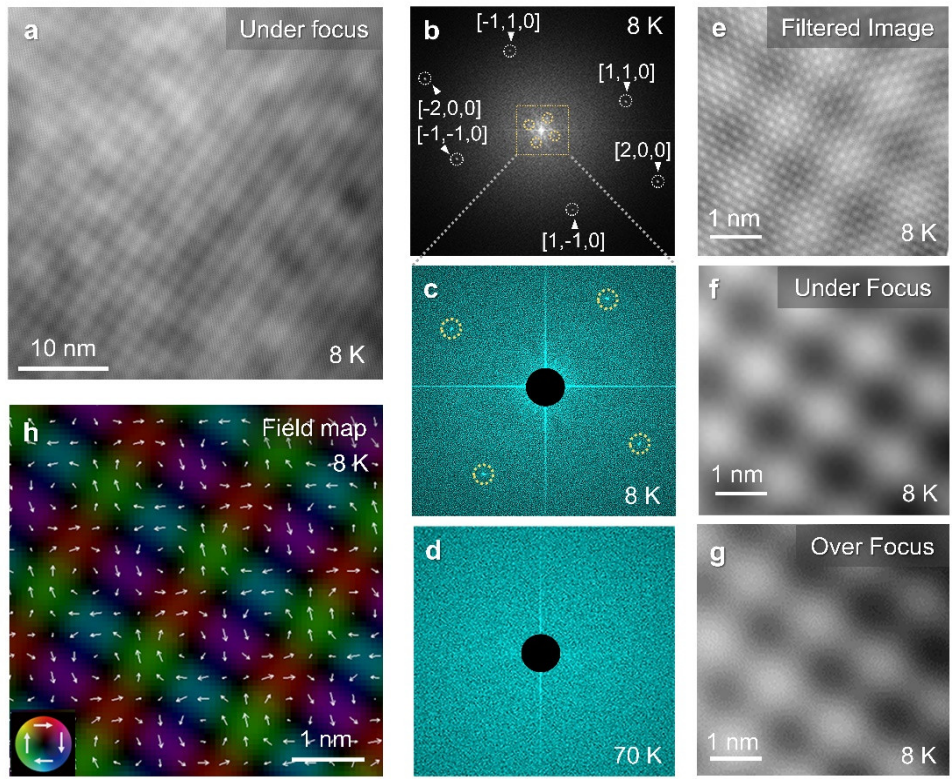

**Fig. 4 | Real space imaging of square SkL in Phase II by Lorentz transmission electron microscopy (L-TEM).** The observation is performed for the (001) plane at $T$ = 8 K and $B$ = 1.95 T (Phase II), unless further specified. $B$ is applied along the out-of-plane [001] direction. **a**, L-TEM image obtained for the under-focused condition. **b**, FFT pattern corresponding to (**a**), including both crystal (dotted white circle) and magnetic (yellow circle) reflections. Here, the [100] reflection is absent due to the extinction rule for space group *I4/mmm*. **c**, Expanded FFT pattern in (**b**) focusing on magnetic reflection. The FFT pattern of an under-focused L-TEM image obtained in the paramagnetic phase ($T$ = 70 K and B = 1.95 T) is also shown in (**d**). **e**, Under-focused L-TEM image in a magnified scale, obtained through both peaks related to crystal and magnetic Bragg reflections in FFT shown in (**b**). **f**,**g**, L-TEM images for under-focused and over-focused conditions, obtained through peaks corresponding to magnetic reflections in FFT patterns of each defocused L-TEM image. **h**, Lateral magnetization distribution obtained from TIE analysis of the L-TEM images. White arrows show the direction of in-plane magnetization, while the background color represents the direction (hue) and relative magnitude (brightness) of the lateral magnetization.